# Fractional Echoes

G. Karras[1], E. Hertz[1], F. Billard[1], B. Lavorel[1], G. Siour[2], J.-M. Hartmann[2], and O. Faucher[1*]

[1]*Laboratoire Interdisciplinaire CARNOT de Bourgogne,
UMR 6303 CNRS-Université Bourgogne Franche-Comté, BP 47870, 21078 Dijon, France and*
[2]*Laboratoire Interuniversitaire des Systèmes Atmosphériques (LISA) CNRS (UMR 7583),
Université Paris Est Créteil, Université Paris Diderot, Institut Pierre-Simon Laplace,
Université Paris Est Créteil, 94010 Créteil Cedex, France*

Erez Gershnabel, Yehiam Prior, and Ilya Sh. Averbukh[†]
*AMOS and Department of Chemical Physics, The Weizmann Institute of Science, Rehovot 76100, Israel*

(Dated: March 3, 2016)

We report the observation of fractional echoes in a double-pulse excited nonlinear system. Unlike standard echoes which appear periodically at delays which are integer multiple of the delay between the two exciting pulses, the fractional echoes appear at rational fractions of this delay. We discuss the mechanism leading to this phenomenon, and provide the first experimental demonstration of fractional echoes by measuring third harmonic generation in a thermal gas of $CO_2$ molecules excited by a pair of femtosecond laser pulses.

PACS numbers: 45.50.-j, 37.10.Vz, 42.50.Md

Many nonlinear systems display the phenomenon of echoes, which is a series of delayed impulsive responses after a pair of short external pulses. These echoes typically show themselves at times $\tau \sim T, 2T, 3T, etc.$ after the end of the excitation, where $T$ is the delay between the stimulating pulses. Echoes are common in many areas of physics, including NMR [1], plasma physics [2–4], nonlinear optics [5], cavity quantum electrodynamics [6, 7], and cold atoms physics [8–11]. Echoes were predicted to occur in proton storage rings [12, 13], and were observed in high energy hadron beam experiments [14, 15]. Echo-enabled generation of short-wavelength radiation in free-electron lasers [16–18] was demonstrated (for a recent review, see [19]).

Recently we reported the existence of orientation/alignment echoes in an ensemble of free classical rotors stimulated by an external impulsive force [20]. We attributed the echo formation to the kick-induced filamentation of the rotational phase space, and demonstrated the effect experimentally in a gas of $CO_2$ molecules excited by a pair of femtosecond laser pulses. The time-dependent mean molecular alignment (defined as $\langle\cos^2(\theta)\rangle(t)$, where $\theta$ is the angle between the molecular axis and laser polarization) was measured via the laser-induced birefringence signal. The excited system exhibited a regular sequence of echo pulses at delays $\tau = T, 2T, 3T, ....$ In addition, our theoretical analysis and numerical simulations [20] also predicted unusual additional recurrences (fractional echoes) at $\tau = (p/q)T$, where $p$ and $q$ are mutually prime numbers. These fractional echoes, however cannot be seen in the alignment signal - they require measurement of higher order moments of the molecular angular distribution.

Here we present the first observation of the phenomenon of fractional echoes measured via third-harmonic generation (THG) in a gas of $CO_2$ molecules at room temperature and pressure of 80 mbar. In what follows we introduce the mechanism of the fractional echoes by using a simple 2D classical model, and then describe our experiments in which the phenomenon of fractional echoes was observed. We also compare the experimental results to the full three-dimensional classical and quantum simulations of the echo formation.

For a nonresonant laser field interacting with a symmetric linear molecule, such as $N_2$, $O_2$ or $CO_2$, the angular-dependent interaction potential is $V(\theta, t) = -(\Delta\alpha/4)E^2(t)\cos^2(\theta)$ [21, 22]. Here $\Delta\alpha$ is the polarizability anisotropy, and $E(t)$ is the envelope of the laser pulse. As is well known, such an interaction leads to the alignment of the molecular axis along the field polarization direction (for reviews on molecular alignment see [23–27]).

We first consider a 2D ensemble of rotors that are initially uniformly dispersed in angle $\theta$, and have a spread in angular velocity $\omega$ with a Gaussian distribution $f(\theta, \omega) \sim \exp[-\omega^2/(2\sigma^2)]$. At time $t = 0$ the ensemble is kicked by a short linearly polarized laser pulse (delta-kick). The phase space probability distribution at time $t$ after the pulse is given by [20]:

$$f(\omega, \theta, t) = \frac{1}{2\pi}\frac{1}{\sqrt{2\pi}\sigma}\exp\left[-\frac{[\omega - \Omega\sin(2\omega t - 2\theta)]^2}{2\sigma^2}\right] \quad (1)$$

Here $\Omega$ is proportional to the time-integrated intensity (total energy) of the pulse.

Figure 1 shows the evolution of the initial distribution with time. After the kick, (Fig. 1a) the shape of the density distribution folds, resulting in transient alignment along the direction $\theta = 0$. On longer time scales, when the alignment signal $\langle\cos^2(\theta)\rangle(t)$ vanishes, the probability density becomes wrinkled, and it develops multiple parallel filaments (Fig. 1b). The number of these filaments grows with time, and in order to keep the phase space volume constant their width is diminishing. Eventually, all the filaments tend to become almost horizontal and uniform in density. As follows from Eq.(1), neighboring filaments are separated in angular velocity by $\pi/t$, where $t$ is the evolution time. Filamentation of the phase space is a phenomenon well known in statistical mechanics of stellar systems [28], and in accelerator physics [29]. Moreover, transient alignment caused by folding of the phase space (see Fig. 1a) has much in common with the bunching effect observed in particle accelerators.

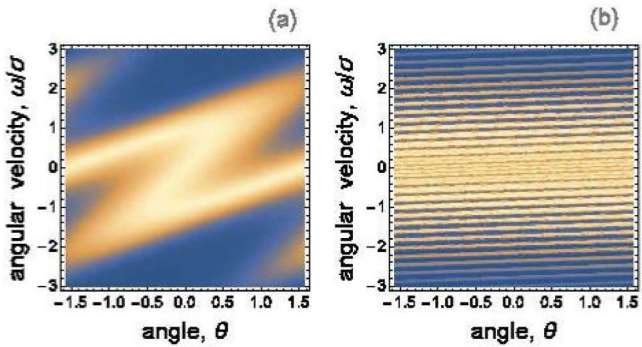

FIG. 1. (color online) Filamentation of the phase space density distribution. $\Omega/\sigma = 1$, (a) $\sigma t = 1$, (b) $\sigma t = 15$.

At $t = T$, the ensemble is subject to a second kick. With time, every filament in Fig. 1b forms the typical folded pattern similar to the description in Fig. 1a. For every angle $\theta$, neighboring filaments are separate in velocity by $\pi/T$ with respect to each other. As a result, after the second kick, most of the time these folded patterns are shifted with respect to each other, which results in a quasi-uniform total angular distribution considered as a function of $\theta$ only. As previously discussed [20], at time $\tau \sim T$ after the second kick, the folded filaments' paths synchronously pile up near $\theta = 0$ due to the "quantization" of the angular velocity of the strips, resulting in an echo in the alignment factor $\langle \cos^2(\theta) \rangle$. Similarly, higher order echoes give rise to transient alignment at delays $2T, 3T, ...$ after the second pulse. An analytic expression for the time-dependent alignment signal given in the Eq.(S-5) in the Supplementary Material of Ref. [20] completely supports these general geometric arguments.

Here, however, we are interested in the highly symmetric structures in phase space, which appear at $\tau = T/2, T/3, ...$ when synchronization of the folded patterns from non-neighboring filaments happens (see Fig. 2 as an example). As discussed in [20], these patterns may be associated with the so called "fractional echoes". These echoes are not seen in a mere alignment signal $\langle \cos^2(\theta) \rangle$, but require measuring higher order observables $\langle \cos(2n\theta) \rangle$ ($n > 1$). The simple 2D model considered here allows for obtaining an analytical expression for the time-dependent mean value of $\langle \cos(2n\theta) \rangle$:

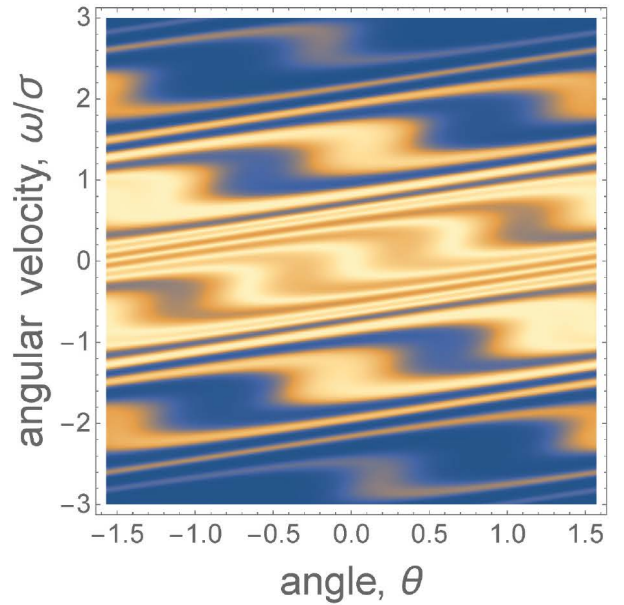

FIG. 2. (color online) Fractional echo in the filamented phase space. $\Omega_1/\sigma = 1$, $\sigma T = 5$, $\Omega_2/\Omega_1 = 1/3$, and $\sigma\tau = 2.293$ - one-half alignment echo is formed.

$$\langle \cos(2n\theta) \rangle(\tau) = \sum_{k=0}^{k=\infty} (-1)^k e^{-2\sigma^2(n\tau - kT)^2} J_{k+n}[2n\Omega_2\tau] J_k[2n\Omega_1(n\tau - kT)], \quad (2)$$

where $J_m(z)$ is the $m$-th order Bessel function of the first kind. Equation (2) presents a sequence of signals localized in time near $\tau = \frac{k}{n}T$ where $k$ is an integer. For $n > 1$, these are the above mentioned "fractional echoes", while $n = 1$ corresponds to the regular alignment echoes. Figure 3 presents the calculated mean values $\langle \cos(2n\theta) \rangle$ versus time after the first kick. It is clearly seen that the initial transient response shortly after the second pulse is followed by a series of fractional echoes at $\tau \sim T/2, T/3$, etc..

All these findings derived in the simplified 2D classical model are confirmed by the results of the fully 3D simulations (both classical and quantum-mechanical) presented below in Figs. 4b, 6b, and 7b. Details of our 3D computational procedures will be given elsewhere. Movie M1 in the Supplementary Material [30] shows the simulated evolution of the angular distribution of an ensemble of 3D classical rotors kicked by two delayed pulses. In addition to common alignment events observed just after the pulses and near times of full echoes, high-order transient symmetric structures can be seen in the angular distribution, which correspond to the fractional echoes (compare with Fig. 2).

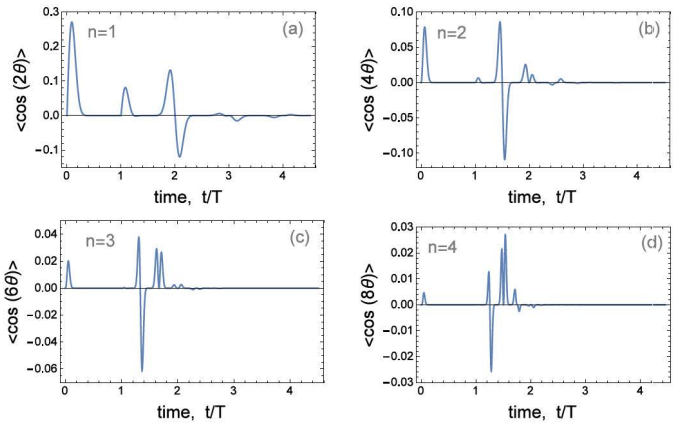

FIG. 3. (color online) Mean values $\langle\cos(2n\theta)\rangle$ versus time after the first kick. $\Omega_1/\sigma = 1$, $\sigma T = 5$, $\Omega_2/\Omega_1 = 1/3$. (a) n=1, (b) n=2, (c) n=3, (c) n=4.

Next, we describe our experiments, where fractional echoes were observed by third harmonic generation in a thermal gas of $CO_2$ molecules excited by a pair of femtosecond laser pulses. In our previous work [20], the observation of the rotational alignment echoes in laser-kicked molecules was performed by using a time-resolved birefringence technique providing a signal $\mathcal{S}_{bir.}$ sensitive to $\langle\cos^2\theta\rangle$

$$\mathcal{S}_{bir.}(\tau) \propto \int_{-\infty}^{\infty} I_{pr}(\tau' - \tau) \left[\langle\cos^2(\theta)\rangle(\tau') - \frac{1}{3}\right]^2 d\tau' \quad (3)$$

with $I_{pr}$ the intensity of the probe pulse. Figure 4a presents the birefringence signal recorded in $CO_2$ molecules by employing the experimental setup detailed in Ref.[20]. Measurements were performed at room temperature for a pressure of 80 mbar. As indicated in the figure legend, the first pump pulse $\mathcal{P}1$ (intensity $I_1 \approx 50$ TW/cm$^2$), the second pump pulse $\mathcal{P}2$ (intensity $I_2 \approx 10$ TW/cm$^2$), and the echoes are colored in red, blue, and yellow, respectively. The first echo appears around $\tau = T = 2.5$ ps, where $T$ is the relative delay between $\mathcal{P}1$ and $\mathcal{P}2$. The second echo (i.e., higher order echo) is detected around $\tau = 2T = 5$ ps, as expected from the above classical arguments. Figure 4b compares the measured signals with the results of the three-dimensional simulations of the birefringence signals, both classical (red curve) and fully quantum mechanical (purple curve). As seen, the classical treatment nicely describes the shape of the first echo, in agreement with the classical origin of this phenomenon. However, the shape of the second echo is already affected by the quantum effects that become important as the delay time $t = T + \tau$ approaches $T_{rev}/4$ - a quarter of the rotational revival period. The intensity dependence of the echo signal with respect to $\mathcal{P}1$ and $\mathcal{P}2$ has been studied in Ref.[20]. The signal observed in Fig. 4a results from the orientational contribution to the optical Kerr effect [31, 32] and therefore only provides information about the observable $\langle\cos^2(\theta)\rangle$. This is the reason why fractional echoes can not be observed with this technique. In order to reveal fractional echoes, the system must be probed through a higher-order nonlinear process sensitive to $\langle\cos^{2m}(\theta)\rangle$, with the integer $m > 1$. Harmonic generation can serve this goal. It has been shown theoretically [33–37] and experimentally [38, 39] that high-order harmonic spectra generated from aligned molecules carry information about higher-order cosine moments. The time dependence of the harmonic amplitude can be generally described by a series of direction cosines products governing the dynamical alignment. The trigonometric terms involved in this expression depend on the molecular system and the harmonic order. For instance, the time dependence of the third-harmonic generation (THG) in aligned $CO_2$ is governed by a combination of observables $\langle\cos^2(\theta)\rangle$ and $\langle\cos^4(\theta)\rangle$ [40].

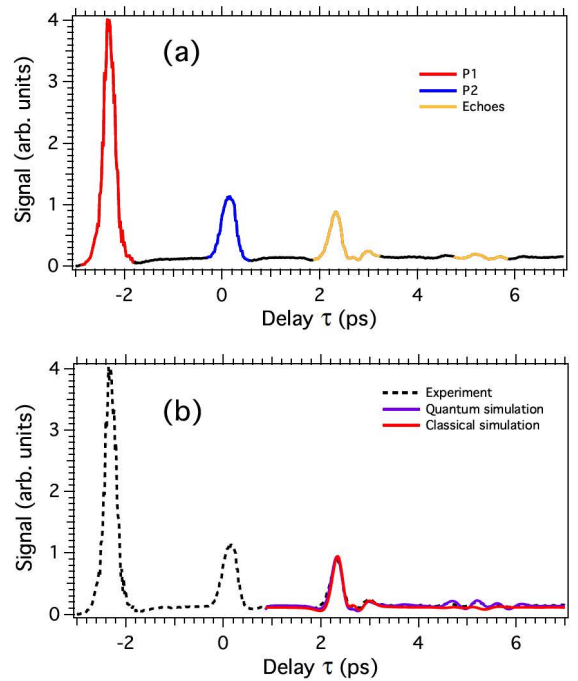

FIG. 4. (color online) (a) Birefringence signal as a function of the delay $\tau$ between $\mathcal{P}2$ and the probe pulse. Only the range corresponding to the quarter rotational period of the $CO_2$ (i.e., $T + \tau < T_{rev}/4 = 10.67$ ps) is shown. The delay $T$ between $\mathcal{P}1$ and $\mathcal{P}2$ is set to 2.5 ps. (b) Comparison between the experiment (dotted black line) and quantum (solid purple line) and classical (solid red line) simulations (see text).

In a recent publication, we showed that a sensitive background-free detection of dynamical alignment could be achieved by generating the THG from a circularly polarized field [41]. As a result of the axial symmetry introduced by the anisotropic alignment, the THG field driven by a circularly polarized pump is elliptically po-larized. The helicity of the harmonic field is governed by the time delay between the alignment pulse and the driving field. The effect is used in the present work in order to observe the fractional echoes in $CO_2$. The signal recorded along each polarization components of the THG field is described in the framework of perturbation theory by the following expression [41]

$$\mathcal{S}_{i_{\text{THG}}}(\tau) \propto \int_{-\infty}^{\infty} I_{\text{pr}}^3 (\tau' - \tau) \left[ \alpha_i \left( \langle \cos^2(\theta)(\tau') \rangle - \frac{1}{3} \right) \right. \\ \left. + \beta_i \left( \langle \cos^4(\theta)(\tau') \rangle - \frac{1}{5} \right) \right]^2 d\tau', \quad (4)$$

where $i = y, z$ are the polarization components of the THG field, with $z$ the direction of alignment of the molecule, and $\alpha_i, \beta_i$ are two parameters that depend on the second-order hyperpolarizability components of the molecule [41]. In order to highlight the presence of the fractional echo, the measurements are conducted by selecting one of the two THG field components [41].

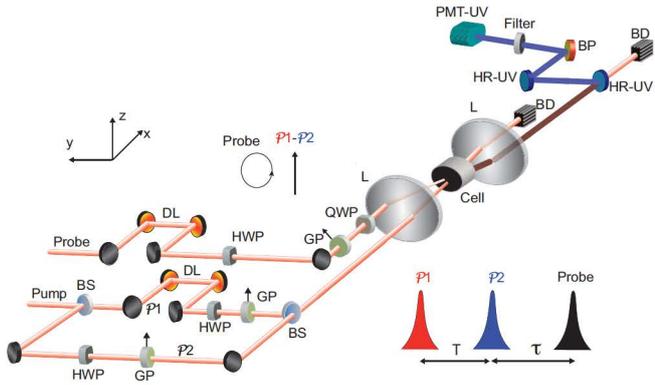

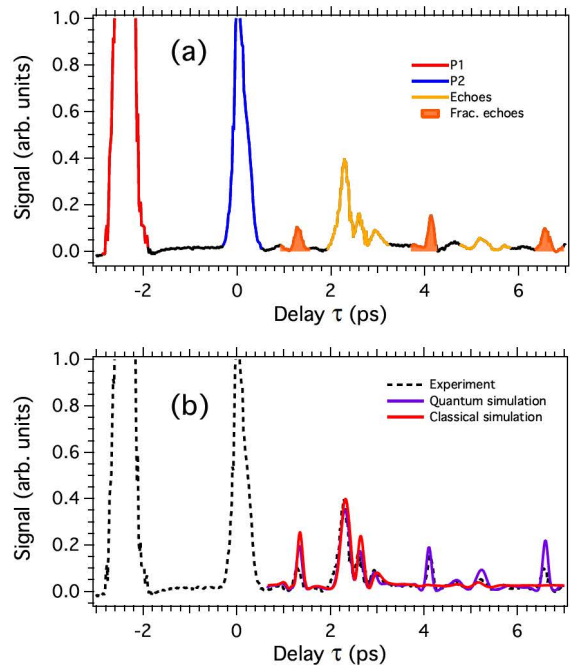

FIG. 5. (color online) Experimental set-up. A: Analyzer, BS: Beam Splitter, GP: Glan Polarizer, HWP: Zero-order half wave-plate, QWP: Zero-order quarter wave-plate PMT: Photo Multiplier Tube, L: Lens, DL: Delay Line, BD: Beam Dumper, HR-UV: High reflectivity dielectric UV mirror and BP: Silica plate set at Brewster's angle. The polarizations of the different pulses along with a relative timing chart are shown in the insets.

The experimental setup for producing THG in aligned molecules is depicted in Fig. 5. The pump and probe IR pulses are produced by a 1 kHz amplified Ti:sapphire laser operating at 800 nm. The pump beam is divided into two pulses denoted as $\mathcal{P}1$ and $\mathcal{P}2$, respectively. The relative delays between the three pulses are controlled by two motorized stages. They allow achieving an accurate adjustment of the delay $T$ between $\mathcal{P}1$ and $\mathcal{P}2$, and $\tau$ between $\mathcal{P}2$ and the probe pulse, respectively. The energy of each pulse is adjusted by combining a half wave-plate with a Glan polarizer. The pump pulses $\mathcal{P}1$ and $\mathcal{P}2$ are linearly polarized, while the polarization of the probe is set circular. The three beams are crossed with a small angle ($\sim 4°$) and focused with a lens (f = 15 cm) inside a static cell filled with $CO_2$ molecules at room temperature. After the cell and a collimating lens the two pump beams are guided to a beam dumper. The THG generated by the probe is passing through two high reflectivity mirrors at 266 nm and a band pass filter for the removal of the fundamental IR frequency component. The polarization sensitive detection is achieved by using a fused silica plate set at Brewster's angle. Finally, the THG signal is recorded using a photo-multiplier tube.

FIG. 6. (color online) (a) Third-harmonic (THG) signal detected by selecting the harmonic field component parallel to the alignment axis $z$. The delay $T$ between $\mathcal{P}1$ and $\mathcal{P}2$ is set to 2.5 ps. (b) Comparison between the experiment (dotted black line) and quantum (solid purple line) and classical (solid red line) simulations.

The THG signal detected along the polarization direction of the pump field is shown in Fig. 6a. The parameters of the pump pulses $\mathcal{P}1$ and $\mathcal{P}2$ are similar to the ones used for the birefringence measurement of Fig. 4a. The energy of the probe is similar to $\mathcal{P}1$. The echo and high-order echoes are observed with the same timing as in the birefringence measurement. However, their structural shapes differ due to the contribution of the observable $\langle \cos^4(\theta) \rangle$ to the THG signal. The main difference between Figs. 4a and 6a is the presence of the fractional echoes in the THG signal (colored in orange) around $\tau = T/2 = 1.25$ ps together with the occurrence of the higher orders around $\tau = 3T/2 = 3.75$ ps and $\tau = 5T/2 = 6.25$ ps, respectively. Figure 6(b) compares experimental signals with the results of our classical and quantum simulations of the THG process. As in the case of the birefringence measurements, the classical and quantum results are practically indistinguishable in the region of the first full echo at $\tau = T = 2.5$ ps, and agree reasonably near the first fractional echo at $\tau = T/2 = 1.25$ ps. Moreover, both curves are in good agreement with the experimental data. However, the classical treatment fails to catch a signal peak near $\tau = 4$ ps, while the quantum modelling reproduces it precisely. Figure 7a shows the THG signal recorded up to 15 ps, i.e. reaching the region near $t \sim T_{\text{rev}}/2$ where the quantum nature of the rotational motion is obviously important. Figure 7b compares it with the results of the fully quantum 3D simulation of the THG process which are in excellent agreement with our measurements. For evolution times shorter than $\sim 4$ psec, the agreement with the 3D classical calculation is also good enough. Multiple spikes in the THG signal seen on the longer time scale result from the interplay between the classical echo effect and the quantum revival phenomenon.

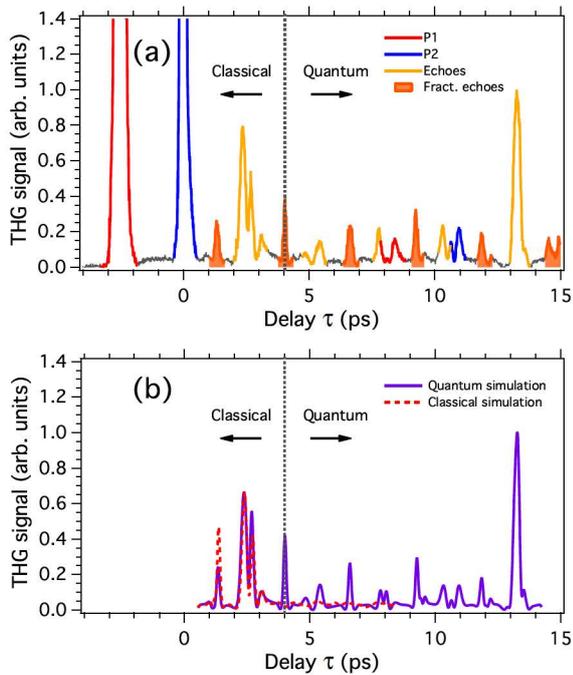

FIG. 7. (color online) (a) Signal detected along the vertical $z$-component of the THG field. The delay $T$ between $\mathcal{P}1$ and $\mathcal{P}2$ is set to 2.6 ps. $I_1 \approx 45\text{TW/cm}^2$, $I_2 \approx 11\text{TW/cm}^2$. (b) Theoretically simulated THG signals: quantum (solid purple line) and classical (dotted red line) simulations.

In conclusion, we observed fractional echoes in the third harmonic signal arising from a thermal gas of $CO_2$ molecules following excitation by a pair of femtosecond laser pulses. These transient responses of the system occur at times which are rational fractions of the delay between the two kicking pulses. We compare the experimental observations to both quantum and classical simulations. The phenomenon of fractional echoes is essentially a classical one, and is well described by classical analysis on short time scales. On longer time scales, the interplay between the echoes and the quantum revivals [42] shows up both in the experiments and in the numerical simulations. The described mechanism of fractional echoes is rather generic, and should be observed in related systems such as echo-enabled harmonic generation in free-electron lasers (EEHG FEL) in which highly efficient generation of short-wavelength radiation is achieved via laser manipulations over the phase space of relativistic electron beams [16–19].

This work was supported by the Conseil Régional de Bourgogne (PARI program), the CNRS, the Labex ACTION program (contract ANR-11-LABX-01-01), and the French National Research Agency (ANR) through the CoConicS program (Contract No. ANR-13-BS08-0013). We also acknowledge financial support by the Israel Science Foundation (Grant No. 601/10), the DFG (Project No. LE 2138/2-1), and the Minerva Foundation. I.A. acknowledges support as the Patricia Elman Bildner Professorial Chair. This research was made possible in part by the historic generosity of the Harold Perlman Family.